\documentclass[11pt,a4paper]{article}
\pdfoutput=1
\usepackage{jcappub}

\title{Constraints on low-mass WIMPs from the EDELWEISS-III dark matter search}

\collaboration{The EDELWEISS Collaboration}

\author[a,\star]{E.~Armengaud}
\author[b,1]{Q.~Arnaud\note{now at Queen's University, Canada}}
\author[b]{C.~Augier}
\author[b]{A.~Beno\^{i}t}
\author[c]{A.~Beno\^{i}t}
\author[d]{L.~Berg\'{e}}
\author[e]{T.~Bergmann}
\author[b]{J.~Billard}
\author[f,g]{J.~Bl\"{u}mer}
\author[a]{T. de~Boissi\`{e}re}
\author[c]{G.~Bres}
\author[d]{A.~Broniatowski}
\author[h]{V.~Brudanin}
\author[c]{P.~Camus}
\author[b]{A.~Cazes}
\author[d]{M.~Chapellier}
\author[b]{F.~Charlieux}
\author[d]{L.~Dumoulin}
\author[g]{K.~Eitel}
\author[h]{D.~Filosofov}
\author[f]{N.~Foerster}
\author[a]{N.~Fourches}
\author[c]{G.~Garde}
\author[b]{J.~Gascon}
\author[a,1]{G.~Gerbier}
\author[d]{A.~Giuliani}
\author[c]{M.~Grollier}
\author[a]{M.~Gros}
\author[g]{L.~Hehn}
\author[a]{S.~Herv\'{e}}
\author[f]{G.~Heuermann}
\author[d]{V.~Humbert}
\author[b]{M. De~J\'{e}sus}
\author[i]{Y.~Jin}
\author[g]{S.~Jokisch}
\author[b]{A.~Juillard}
\author[b,f]{C.~K\'{e}f\'{e}lian}
\author[e]{M.~Kleifges}
\author[g]{V.~Kozlov}
\author[j]{H.~Kraus}
\author[k]{V. A.~Kudryavtsev}
\author[d]{H.~Le-Sueur}
\author[j]{J.~Lin}
\author[d]{M.~Mancuso}
\author[d]{S.~Marnieros}
\author[e]{A.~Menshikov}
\author[a]{X.-F.~Navick}
\author[a]{C.~Nones}
\author[d]{E.~Olivieri}
\author[l]{P.~Pari}
\author[a]{B.~Paul}
\author[d,2]{M.-C.~Piro\note{Now at Rensselaer Polytechnic Institute, Troy, USA}}
\author[d]{D.~V.~Poda}
\author[b]{E.~Queguiner}
\author[k]{M.~Robinson}
\author[c]{H.~Rodenas}
\author[h]{S.~Rozov}
\author[b]{V.~Sanglard}
\author[g,3]{B.~Schmidt\note{Now at Lawrence Berkeley National Laboratory, Berkeley, USA}}
\author[f]{S.~Scorza}
\author[g]{B.~Siebenborn}
\author[e]{D.~Tcherniakhovski}
\author[b]{L.~Vagneron}
\author[e]{M.~Weber}
\author[h]{E.~Yakushev}
\author[j]{X.~Zhang}

\affiliation[a]{CEA Saclay, DSM/IRFU, 91191 Gif-sur-Yvette Cedex, France}
\affiliation[b]{Institut de Physique Nucl\'{e}aire de Lyon-UCBL, IN2P3-CNRS, 4 rue Enrico Fermi, 69622 Villeurbanne Cedex, France}
\affiliation[c]{Institut N\'{e}el, CNRS/UJF, 25 rue des Martyrs, BP 166, 38042 Grenoble, France}
\affiliation[d]{CSNSM, Univ. Paris-Sud, CNRS/IN2P3, Universit\'{e} Paris-Saclay, 91405 Orsay, France}
\affiliation[e]{Karlsruher Institut f\"{u}r Technologie, Institut f\"{u}r Prozessdatenverarbeitung und Elektronik, Postfach 3640, 76021 
Karlsruhe, Germany}
\affiliation[f]{Karlsruher Institut f\"{u}r Technologie, Institut f\"{u}r Experimentelle Kernphysik, Gaedestr. 1, 76128 Karlsruhe, Germany}
\affiliation[g]{Karlsruher Institut f\"{u}r Technologie, Institut f\"{u}r Kernphysik, Postfach 3640, 76021 Karlsruhe, Germany}
\affiliation[h]{JINR, Laboratory of Nuclear Problems, Joliot-Curie 6, 141980 Dubna, Moscow Region, Russian Federation}
\affiliation[i]{Laboratoire de Photonique et de Nanostructures, CNRS, Route de Nozay, 91460 Marcoussis, France}
\affiliation[j]{University of Oxford, Department of Physics, Keble Road, Oxford OX1 3RH, UK}
\affiliation[k]{University of Sheffield, Department of Physics and Astronomy, Sheffield, S3 7RH, UK}
\affiliation[l]{CEA Saclay, DSM/IRAMIS, 91191 Gif-sur-Yvette Cedex, France}
\affiliation[\star]{corresponding author}

\emailAdd{eric.armengaud@cea.fr} 

\abstract{We present the results of a search for elastic scattering from galactic dark matter in the form of Weakly Interacting Massive Particles (WIMPs) in the $4-30\, {\rm GeV}/c^2$ mass range. We make use of a 582~kg-day fiducial exposure from an array of 800~g Germanium bolometers equipped with a set of interleaved electrodes with full surface coverage. We searched specifically for $\sim 2.5-20$~keV nuclear recoils inside the detector fiducial volume. As an illustration the number of observed events in the search for 5 (resp.~20)~${\rm GeV}/c^2$ WIMPs are 9 (resp.~4), compared to an expected background of 6.1 (resp.~1.4). A 90\% CL limit of $4.3\times 10^{-40}$~cm$^2$ (resp. $9.4\times 10^{-44}$~cm$^2$) is set on the spin-independent WIMP-nucleon scattering cross-section for 5 (resp. 20)~${\rm GeV}/c^2$ WIMPs. This result represents a 41-fold improvement with respect to the previous EDELWEISS-II low-mass WIMP search for 7~${\rm GeV}/c^2$ WIMPs. The derived constraint is in tension with hints of WIMP signals from some recent experiments, thus confirming results obtained with different detection techniques.}

\keywords{dark matter detectors, dark matter experiments}

\begin{document}
\maketitle

\section{Introduction}

A vast array of observations from galactic to the largest observable scales in the Universe is currently best interpreted within the $\Lambda$CDM framework~\cite{planck}, which assumes the existence of non-baryonic dark matter as one of its foundations~\cite{bertone}. Since the dark matter properties are very poorly constrained as of now, many models have been proposed to describe its nature. Weakly Interacting Massive Particles (WIMPs) constitute a generic candidate with a mass in the GeV to TeV range. WIMPs from our galactic halo can scatter off atomic nuclei and generate detectable keV-scale nuclear recoils~\cite{goodman}.

In this article, we will focus on the direct detection of WIMPs with mass $M_X$ ranging from 4 to 30~${\,\rm GeV}/c^2$. Thermal relics in this mass range are somewhat disfavored both from constraints set by Fermi-LAT searches for annihilation signals in dwarf galaxies~\cite{fermi}, and from the impact that WIMP annihilation would leave on the cosmic microwave background anisotropies~\cite{planck-2015}. However, many scenarios have been proposed where the evolution of the "dark" and "visible" sectors in the early Universe are such that the relic dark matter number density is naturally close to the baryon density~\cite{kaplan,falkowski,petraki}. The current measurements of the corresponding mass densities, $\Omega_c$ and $\Omega_b$, then imply that $M_X\sim 5 \,{\rm GeV}/c^2$, which we will use as a benchmark WIMP mass for this search.

In addition, several dark matter search experiments~\cite{dama,cogent,cresst,cdms-si} using different target nuclei and experimental approaches found hints of WIMP dark matter whose mass could be in the 6 to $50\, {\rm GeV}/c^2$ range, and WIMP-nucleon elastic cross-section in the $10^{-40} - 10^{-42}$~cm$^2$ range. The consistency between these hints was debated~\cite{kelko,kopp}. More importantly they are already severely constrained by other experimental searches~\cite{lux,scdms}. The aim of this article is also  to provide an independent check of these former results.

To search for low-energy WIMP-induced nuclear recoils, detectors with both low detection threshold and a very low background are needed. The EDELWEISS experiment operates high-purity Germanium crystals in a cryogenic low-background environment, inside the Laboratoire Souterrain de Modane (LSM). For each detector, the combined measurement of heat and ionisation signals enables the discrimination of a potential WIMP-induced signal against several radioactive backgrounds. The latest generation of EDELWEISS detectors, equipped with a set of Fully InterDigitized electrodes (FID detectors~\cite{fid-gamma}), provides the possibility to select interactions taking place inside a large fiducial fraction of each detector. During the third stage of the EDELWEISS experiment, a ten-month-long run in low-background conditions was carried out with an array of twenty-four $\sim 800$-g FID detectors. We exploit here the data from eight of those detectors to carry out a blind search for low-mass WIMPs. In the following sections, we will first present in some details the detector performance and backgrounds of relevance for this search. We will then describe the WIMP search itself, which makes use of a Boosted Decision Tree (BDT) algorithm dedicated to background rejection. Finally, the lack of a significant excess of WIMP-like events over the predicted backgrounds is exploited to derive a set of limits on the WIMP-nucleon cross-sections. This result extends the sensitivity of the previous low-mass WIMP search carried out with EDELWEISS-II detectors~\cite{edw2-lowmass}, thanks to a higher exposure, lower thresholds and better background rejection.

\section{Detector performance and backgrounds}

\subsection{Overview of the experimental setup}

\begin{figure}[!ht]
\centering
\includegraphics[width=0.9\textwidth]{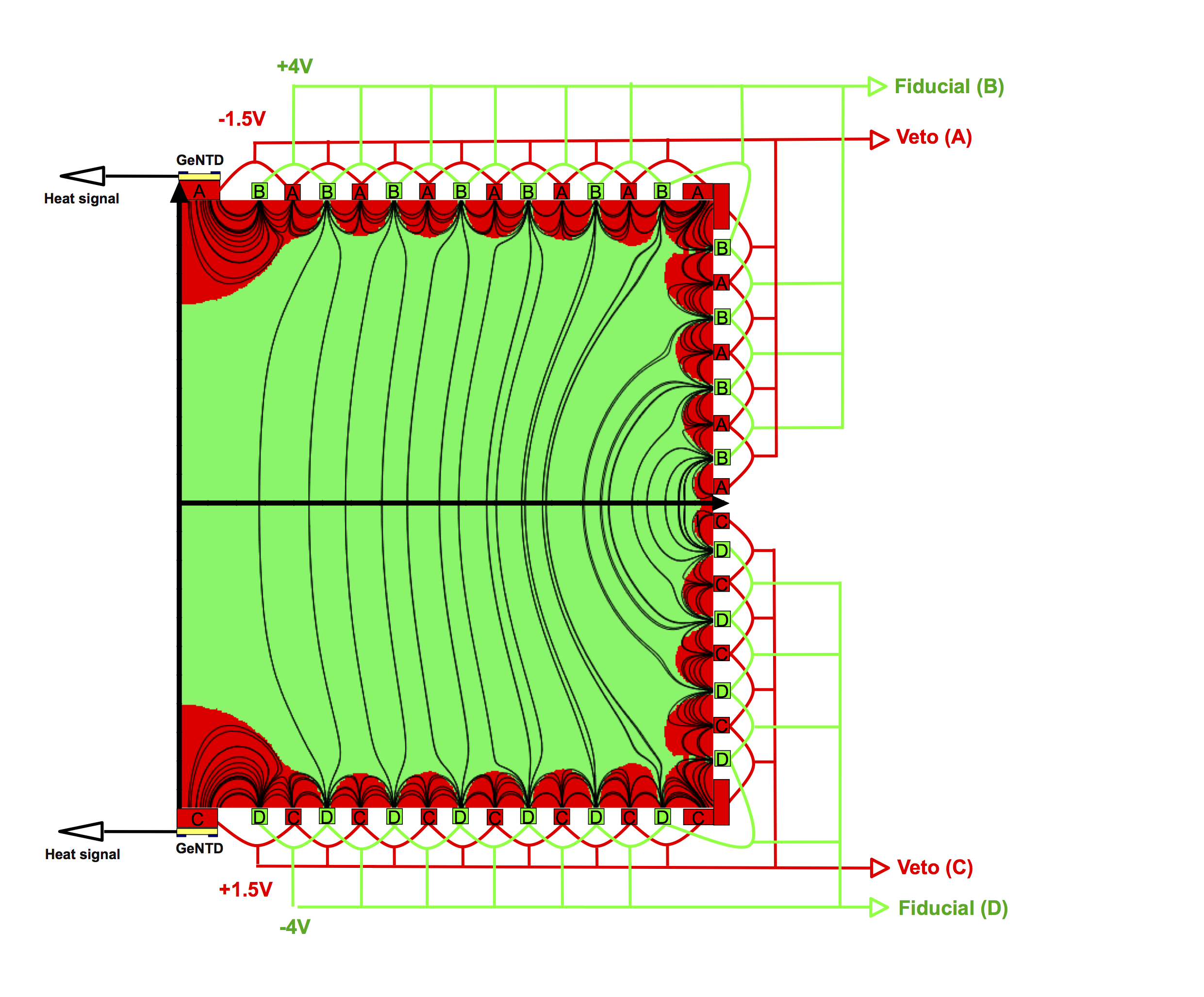}
\caption{Cross-sectional view of an 800-g FID detector. The fiducial and veto electrodes are represented with their corresponding wirings. Electric field lines are shown inside the detector, as derived from a numerical calculation. The associated fiducial volume is represented in green, while charge depositions taking place inside the red regions are tagged by signals on a veto electrode.}
\label{fig:detector}
\end{figure}

The detectors used in this search consist of 820 to 890~g cylindric high-purity Ge monocrystals with an appropriate surface treatment~\cite{edw-etching}. Fig.~\ref{fig:detector} illustrates the operation principle of these detectors. Interleaved ring electrodes were evaporated over both their planar and side surfaces. Under nominal operation, these electrodes are polarised alternatively at $+4$ and $-1.5$~V on the top side, and opposite-sign voltages on the bottom. All electrodes at a given voltage are wired together, so that there are four charge readout channels per detector. The resulting electric field inside the crystal causes charges created far from the surface to generate a time-integrated signal only on the so-called fiducial electrodes polarized at $\pm 4$~V. This defines the so-called fiducial volume, whose mass is on average 625~g. The fiducial volume results from the electric field geometry and charge propagation effects, and does not depend on the mechanism generating the initial charge deposition. Charges created close to the surface will generate signals both on the fiducial and the veto electrodes (polarized at $\mp 1.5$~V). For each detector, two neutron transmutation doped (NTD) thermistors are glued on the top and bottom surfaces. When operated at low temperature, they provide a measurement of the total energy deposited in the detector after thermalisation, in an identical way as for EDELWEISS-I~\cite{edw1} and EDELWEISS-II~\cite{edw2}.

These detectors differ from the ID bolometers used in the previous phase of the experiment~\cite{id,edw2} by their large mass, and by being fully covered with interleaved electrodes, hence the name "FID". However, their working principle is similar. For each recorded interaction, the measurement of charge topology enables the rejection of surface interactions such as those generated by $^{210}$Pb subsequent decays at the detector surface and immediate surroundings. The corresponding rejection factor for 800-g FID detectors is better than $4\times 10^{-5}$ (90\% CL) above 15~keV recoil energies, as derived from the exposure of two detectors to $^{210}$Pb sources~\cite{fid-beta}. After fiducial selection, the combined heat and fiducial ionisation measurements provide a way to distinguish electron recoils (ER) and nuclear recoils (NR), since they have different ionisation yields. In~\cite{fid-gamma}, it was shown using $^{133}$Ba sources that the fraction of electron recoils present in a region with 90\% acceptance for nuclear recoils is less than $6\times 10^{-6}$ (90\% CL). No sign of a degradation of this rejection performance as a function of energy is observed, except for the case when the signal amplitudes are similar to their baseline fluctuations.

The data used in this search were taken in the EDELWEISS cryostat at LSM during a ten-month-long run between July 2014 and April 2015, where the cryostat was kept at a stabilized 18~mK temperature during all the data-taking periods. The experimental setup is presented in detail in~\cite{edw-bg,edw-setup}. The WIMP target is an array of twenty-four FID detectors in a compact geometry of six four-unit towers, supported by a high-purity copper structure. Each detector is mounted inside a high-purity copper casing, and is held inside this casing with small teflon holders. The cryostat is a reverse-geometry dilution fridge, inside a set of radiopure copper thermal screens at 1~K, 4.2~K, 40~K and 100~K. 

Gamma radiation is attenuated by an 18~cm thick layer of modern lead lined with a 2~cm inner layer of Roman lead, together with a 14~cm Roman lead plate cooled at 1~K to shield the detectors from the cold electronics. The resulting background measured in the fiducial volume of the detectors in the $20-200$~keV range is 70 counts per kg-day~\cite{edw-bg-lrt}. Polyethylene shields consist of an outer 50~cm layer, an additional 5~cm layer and a 10~cm plate screening the detectors from the cold electronics. They were designed to reduce the single neutron scatter rate in the detectors to $3\times 10^{-4}$~events per kg-day~\cite{edw-bg}. The entire outer polyethylene shield is covered by plastic scintillators (98\% geometrical efficiency for throughgoing muons), read out in coincidence with the bolometers to reject muon-induced events~\cite{edw-veto}.

The cold and warm electronics are described in~\cite{edw-setup,edw-elec,edw-crate}. The main difference with respect to the electronics chain used previously by EDELWEISS (eg.~\cite{edw2}) concerns the ionisation channels. The first stage of amplification consists of a field-effect transistor follower with no bias and feedback resistors to avoid their thermal noises. Based on the ADC output readout sampled at 100~kHz, the gate voltage is periodically adjusted through a capacitance by a FPGA-based software controlling a DAQ. As a consequence, the response of the system to a charge deposit is a step function. The integration time for the charge signal is then only limited by pile-up. With the adopted integration time of 1~sec, we obtained baseline resolutions a factor two better on average than those achieved in~\cite{edw2}. The heat signals are described and modelled in~\cite{edw-thermal}: their rise time of the order of 5~ms is followed by three decays of $\sim10$, 100 and 1000~ms.

\subsection{Event recording and selection}

For nuclear recoils, the observed signal-to-noise ratio is better for the heat signal than for the ionisation one. The decision to store data to disk is therefore triggered by comparing the amplitude of the filtered heat channels to a reference online threshold. The filter consists of a Finite-Impulse-Response Butterworth high-pass filter followed by a convolution with a predefined pulse template. The stored data consist of the raw traces of all channels of the corresponding detector and its neighbors. In WIMP search runs, the online threshold is automatically adjusted by an algorithm designed to lower the threshold as much as possible without exceeding a trigger rate of 50~mHz, i.e. ten times the rate of physical pulses.

From both the heat baseline resolution measurement at a given time $\sigma(t)$ and the value of instantaneous trigger level $E_{\rm thr}(t)$, we model the trigger efficiency as a function of heat energy $E$ with the function: 

$$\epsilon(E,t) = \frac{1}{2} \left[1+{\rm Erf}\left(\frac{E-E_{\rm thr}(t)}{\sqrt{2}\,\sigma(t)}\right)\right] $$

\begin{figure}[!ht]
\centering
\includegraphics[width=0.9\textwidth]{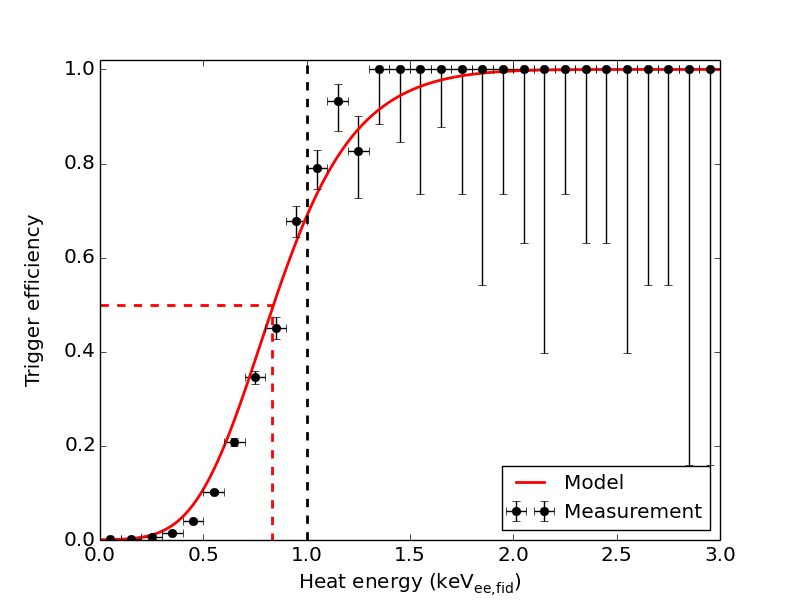}
\caption{Red curve: time-averaged trigger efficiency function as computed in our model (see text) for the WIMP-search selected data of the detector FID825. Data points: corresponding measurement based on events in coincidence with another detector, within the same time period. The dashed lines illustrate the 50\% trigger efficiency at 0.83~keV (red) and the 1.0~keV analysis threshold for this detector (black). The energy scale corresponds to fiducial electron recoils (see text).}
\label{fig:trigger}
\end{figure}

\noindent The agreement between this model and measurements using coincidences either in WIMP search runs or with neutron calibration data is good, as illustrated in Fig.~\ref{fig:trigger}.

The reconstruction of each event is done simultaneously from the different channels that are recorded. The signal amplitudes and their timing are fitted using template shapes. The amplitudes are adjusted under the constraint of the simultaneity of all pulses in the event, at the time given by the most significant one in terms of signal-to-noise ratio. The heat amplitudes are fit in the frequency domain, using the noise power spectrum extracted from the data between triggers and updated at an hourly basis. For the ionisation channels, to avoid windowing biases due to the pulses extending far outside the recorded samples, fits are performed in the time domain, after applying a numerical filter. Cross-talk corrections and energy calibrations are carried out for each detector using events recorded in the presence of two $^{133}$Ba gamma sources which were regularly used during the run. All the energies are calibrated in terms of fiducial electron recoils (keV$_{\rm ee,fid}$). The non-linearities of the ionisation response are negligible below a few hundreds of keV. On the other hand, a correction is applied to the heat energy scale down to low energy to take into account the measured non-linearities.

For a given detector, the heat signals of both NTDs are combined into one using weights that optimize the resolution. The fiducial ionisation signal is defined as the average of the signals on the two electrodes biased at $\pm 4$~V~\cite{edw-trapping,thesis-quentin}, while the total ionisation is defined as the average of all four electrode signals.

Appart from a few detectors with specific issues such as a nonfunctional channel, the main difference between detectors in terms of  performance is related to variations in their heat noise, which drives their online threshold. In order to optimize the sensitivity to WIMP masses in the 4 to $30 \,{\rm GeV}/c^2$ range, we decided to use eight detectors with the best online threshold distributions. For each detector, we selected time periods when the online threshold is less than 1.5~keV$_{\rm ee,fid}$. In addition, we rejected the small fraction of time when the combined heat baseline FWHM was larger than 1~keV$_{\rm ee,fid}$, the fiducial ionisation FWHM was larger than 0.7~keV$_{\rm ee,fid}$, or one of the veto electrode FWHM was larger than 1.5~keV$_{\rm ee,fid}$. This results in a total of 927 live detector-days for WIMP searches, including the effects of various DAQ-related deadtimes. The fiducial fraction was measured for each detector using the triplet of cosmic activation lines associated to $^{65}{\rm Zn}$, $^{68}{\rm Ga}$ and $^{68}{\rm Ge}$, at 9.0, 9.7 and 10.4~keV, in the same way as in~\cite{edw2}: the electron recoils associated to those lines are distributed homogeneously inside the detector volume. The total intensity of these lines was measured using both heat and ionisation signals. The line intensity inside the fiducial volume is then estimated after applying a cut of $<2.58\,\sigma$ of the baseline on the signals from veto electrodes. The fiducial fraction is the ratio of the fiducial over total line intensities. In the energy range of interest, its variation with energy is a small effect, taken into account later in the analysis. The resulting exposure inside the fiducial volume is 582 kg-days. Table~\ref{tbl:boloperf} summarizes the most relevant performances of individual detectors.

\begin{table}[!ht]
\centering
\setlength{\tabcolsep}{0.2cm}
\begin{tabular}{|l|c|c|c|c|c|}
\hline
Detector \rule{0in}{3ex} & Heat & Ionisation & Online & Analysis &  Exposure \\
 & FWHM & FWHM & threshold & threshold & (kg-days) \\[0.15cm]
\hline \hline
FID824 \rule{0in}{3ex}& 0.30 & 0.53 & 0.62 & 1.0 & 59  \\[0.15cm]
FID825 & 0.47 & 0.45 & 0.83 & 1.0 & 76 \\[0.15cm]
FID827 & 0.40 & 0.50 & 0.79 & 1.0 & 78 \\[0.15cm]
FID837 & 0.39 & 0.52 & 0.91 & 1.5 & 73 \\[0.15cm]
FID838 & 0.42 & 0.51 & 0.83 & 1.0 & 75 \\[0.15cm]
FID839 & 0.59 & 0.54 & 1.15 & 1.5 & 75 \\[0.15cm]
FID841 & 0.52 & 0.47 & 0.95 & 1.5 & 86 \\[0.15cm]
FID842 & 0.57 & 0.62 & 1.15 & 1.5 & 59 \\[0.15cm]
\hline
\end{tabular}
\caption{Median key detector performance for the data selected in this WIMP search. All energies are in keV$_{\rm ee,fid}$. The ionisation baseline fluctuations (FWHM) correspond to the combined fiducial ionisation signal. The quoted online thresholds are 50\% efficiency values. The analysis thresholds are described in Section~\ref{sec:wimpsearch}.}
\label{tbl:boloperf}
\end{table}

In order to select only well-reconstructed pulses and reject pile-ups, we applied cuts on the $\chi^2$ resulting from individual trace adjustments. The efficiency of these cuts is measured using either the $9.0-10.4$~keV triplet, or unbiased traces. Both methods are in good agreement. The average efficiency loss is 9\%, mostly driven by the heat channels. To reject interactions taking place inside a single NTD sensor, an additional cut is applied on the difference between the two heat channels of a given detector, resulting in an efficiency loss smaller than 1\%.

Finally, after the full offline reconstruction, events that triggered a single detector, with less than $5\sigma$ veto signal, a reconstructed ionisation yield between 0 and 0.5, and a reconstructed heat energy between 0 and 200 keV, were blinded until all analysis parameters for this WIMP search were frozen.

\subsection{Backgrounds}\label{sec:bg}

We now turn to the description of all expected backgrounds encountered in or close to the region of parameter space where the WIMP signal is expected, i.e. for energies smaller than $\sim 20$~keV$_{\rm ee,fid}$. For each background component, a model was derived before unblinding, using data sidebands that do not overlap with the WIMP signal region.

A nuclear recoil induced by the residual neutron background is indistinguishable from a WIMP-related signal, in the case when a single fiducial interaction is recorded and the recoil energy is similar to that expected from WIMPs. Simulations show that the muon-induced single NR background for the selected data is $0.45\pm0.03\,{\rm (stat)}\,^{+0.14}_{-0.09}$ (syst) events, in good agreement with the number of detected muon-induced events. Given the high efficiency of the muon veto $>93$~\%, the resulting background after rejection of events in coincidence with muon veto hits is expected to be $<0.04$ counts at 90\% CL for the selected data set~\cite{thesis-cecile}. 

Multiple scatters in the NR region were monitored during the WIMP search run using all available detectors. Nine nuclear recoils in the $10-100$~keV range were observed in coincidence with another detector, but not with the muon veto, in a 1309 kg-days exposure. Simulations involving all known sources of radiogenic neutrons~\cite{edw-bg-lrt} cannot reproduce such a number, thus hinting at another, yet unknown source of neutrons. Simulations show that the spectral shape of the neutron-induced nuclear recoil distribution does not depend significantly on the location of the neutron source, and can be approximated in the $2-20$~keV energy range by a double exponential law. Moreover, simulations indicate that the ratio of single over multiple events in the $10-100$~keV range is 0.45 for our detector configuration, with little dependence on the exact location of the neutron source. We therefore built an empirical model for our fiducial, single radiogenic neutron-induced NRs which consists in a double-exponential energy distribution, normalised according to the observed multiple events and the expected single over multiple count ratio. The systematics on the intensity of this background is 45\%, mostly driven by the statistics of observed multiple scatters ($1/\sqrt{9}$) and a 30\% uncertainty on the single over multiple ratio derived from simulations.

\begin{figure}[!ht]
\centering
\includegraphics[height=0.4\textheight]{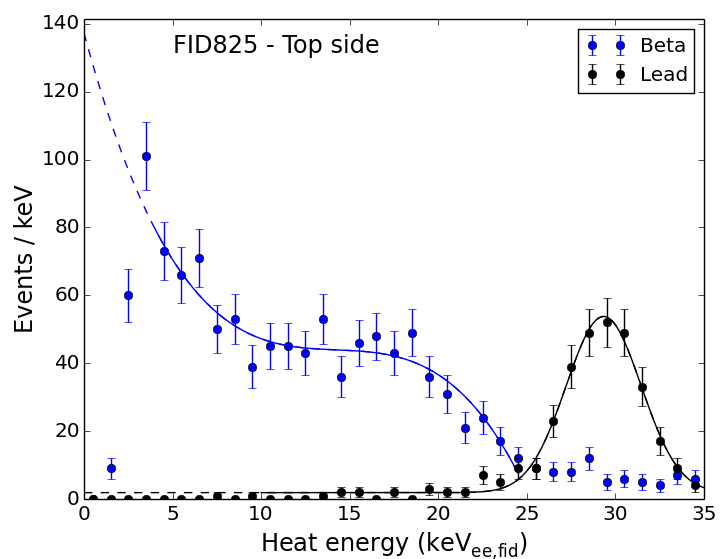}
\includegraphics[height=0.4\textheight]{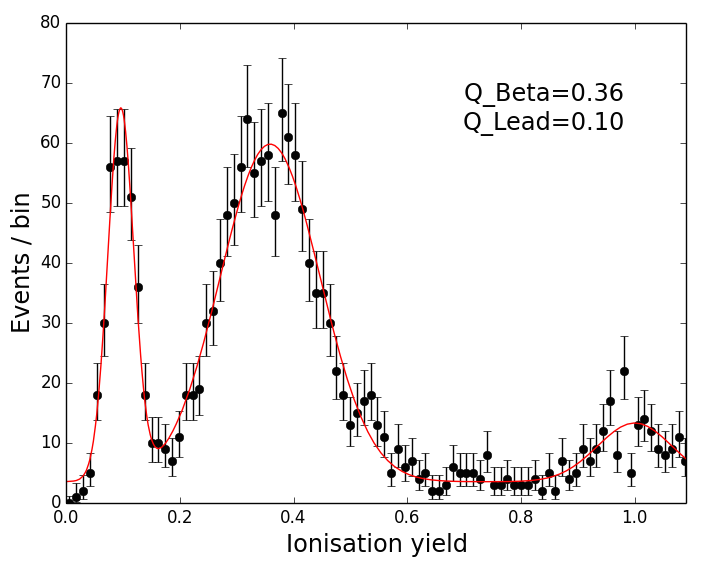}
\caption{Top: heat energy spectra for $\beta$ and lead events on the top side of FID825 for all WIMP-search data. Continuous (resp. dashed) lines represent the fit (resp. extrapolation) to the model as described in the text. Bottom: corresponding ionisation yield distribution, for all top side events with heat energy $<40$~keV$_{\rm ee,fid}$. The ionisation yield is defined here such that surface gamma events have an average yield equal to one.}
\label{fig:betas}
\end{figure}

Radioactive contamination associated for example to radon daughters at the surface of the detectors, or the surface of materials in direct view, generate a significant amount of low-energy beta radiation and lead recoils recorded as surface events. The associated charge deposition on the fiducial and veto electrodes of a given detector side has typical ionisation yields of 0.4 for betas, and 0 or 0.1 for lead recoils (depending on whether the recoil hits an electrode or the surface between electrodes). It is difficult to predict the energy spectrum of these events from ab-initio simulations due to large uncertainties, e.g. in the implantation depth of radioactive contaminations or in the charge collection of very low energy surface betas. Instead, we measured directly the beta and lead ionisation yields and energy spectra for each detector side by selecting events according to their ionisation topology. We define top (resp. bottom) side events by requiring signals more than $5\,\sigma$ (baseline noise) on the electrodes associated to the relevant detector side. Example distributions are shown on Fig.~\ref{fig:betas} in the case of the top side for FID825. For betas, the spectra were fitted to spline functions down to 4~keV$_{\rm ee,fid}$ heat signal, and then extrapolated down to 0~keV. For lead recoils with non-zero ionisation yield, we used the sum of a gaussian and flat distribution to fit their spectra and extrapolate them down to 0~keV. Using these parameterizations, the average event rates in the 927 detector-days exposure used for WIMP search are 20.1 betas per kg-day and 4.5 lead recoils with non-zero ionisation yield per kg-day. Lead recoils with null ionisation yield cannot be distinguished from the heat-only events that we describe in the last paragraph of this Section.
The dominant uncertainty on this surface event background parameterization comes from the low-energy extrapolation of the $\beta$-induced spectrum. Simulations indicate a strong model dependence in the rise at low energy, hence we conservatively model the related systematics by either assuming that the spectrum is flat below 4~keV$_{\rm ee,fid}$, or that its slope is doubled with respect to the reference extrapolation. These extreme assumptions propagate into a $\pm 30$~\% associated systematic error in the predicted numbers after all cuts described in Sect.~\ref{sec:wimpsearch} for a $M_X=5\,{\rm GeV}/c^2$ WIMP search. Even by assuming such a large systematics, Table~\ref{tbl:unblinding} shows that the impact of this uncertainty is small on the overall background uncertainty budget.

Apart from this surface radioactivity, the low-energy electron recoil background consists in a set of cosmogenic activation lines, together with a continuous component associated to Compton scattering and to tritium beta decay~\cite{edw-tritium}. The cosmogenic lines of relevance here are the aforementioned K-shell triplet in the range $9.0-10.4$~keV, the corresponding L-shell lines, and a set of other lines with lower intensities in the $5.0-7.7$~keV range. 
Depending on their location inside the detector and on their multiplicity, these interactions can generate different ionisation signal topologies: purely fiducial events with charges on both fiducial electrodes, near-surface events with charges on the fiducial and veto electrodes of a given side, and triple events with charges on both fiducial electrodes and one veto. For each of these event categories, a model electron recoil spectrum is derived using the measured energy distribution in the $3-12$~keV$_{\rm ee,fid}$ range. It is extrapolated to lower energy under the assumption that the continuous component is flat below 8~keV, and that the L/K electron capture intensity ratio is 0.11~\cite{bahcall}. For our WIMP-search data, the average low-energy continuous ER fiducial background is 0.18 counts per kg-day per keV$_{\rm ee,fid}$. The average background expected from L-shell lines is 0.79 counts per kg-day. There are several sources of uncertainty for this fiducial ER background, whose impacts are similar: the expected L/K ratio, statistical fit errors, systematics in our trigger efficiency model and assumptions on the shape of the continuous component. Simulations show that the total systematic error on the number of fiducial ER events in the energy range of interest for a 5~${\rm GeV}/c^2$ WIMP search is $\pm 16$~\%.

An additional source of background consists in a large number of "heat-only" events, with a significant heat signal but no associated ionisation except  gaussian noise. The shape of the energy distribution for these events is observed to be independent of the detector and time, and is well described by a double-exponential distribution, dominated by a term proportionnal to $e^{-E/3\,{\rm keV}_{\rm ee,fid}}$. Below $1-2$~keV$_{\rm ee,fid}$ (depending on the heat noise), this population overlaps with that of pure noise events, which are recorded when the acquisition triggers on a baseline fluctuation. The rate of heat-only events is very variable as a function of both detector and time. Above 3~keV$_{\rm ee,fid}$, it varies from 1 up to 100 events per day and per detector. The origin of heat-only events is still under investigation, but they are potentially associated with cracks taking place at the level of individual detector's holders, in a similar way to that described in~\cite{cresst-cracks}. To model the distributions in rate and in amplitude of these events in each detector, we used the observed distribution of events with negative total ionisation signal. Using WIMP signal simulations as described in Sect..~\ref{sec:wimpsearch}, it was checked that above 1~keV$_{\rm ee,fid}$, this sample does not overlap with potential WIMP signals. While the origin of these heat-only events is still not fully determined, the systematics on the data-driven background model is well controlled thanks to the large sideband data available. We found that the strongest source of uncertainty comes from a small cross-talk between heat and fiducial ionisation channels, at the level of $0.0-0.7$~\% depending on the detector, which biases the observed ionisation distribution of these events towards positive values, with an energy-dependent bias. In the worst case, after all cuts described in Sect.~\ref{sec:wimpsearch} in a search for a 5~${\rm GeV}/c^2$ WIMP, simulations show that the heat-only background could be larger by 17\%.

\section{WIMP search}

\subsection{Search procedure}\label{sec:wimpsearch}

The WIMP search procedure consists in a set of cuts determined using simulations of our background model and the expected signal, for each WIMP mass in the interval $4-30\,{\rm GeV}/c^2$. After applying the cuts, a simple event counting is performed without background subtraction. All the cuts were defined before unblinding the data in the region of interest.

First of all, analysis thresholds on heat energies are chosen such that the corresponding average trigger efficiency is at least 70\% at the threshold level, in order to reduce the impact on the uncertainties in the modeling of the trigger. For four of the detectors this analysis threshold is 1~keV$_{\rm ee,fid}$ and for the four others it is 1.5~keV$_{\rm ee,fid}$. In addition, the following pre-selection fully defines the experimental region of interest (RoI) on which the WIMP search is focused: a positive total ionisation signal, a signal of less than $5\,\sigma$ of baseline fluctuations on either veto electrodes, heat signal smaller than 12~keV$_{\rm ee,fid}$ and ionisation less than 8~keV$_{\rm ee,fid}$.

\begin{figure}[!ht]
\centering
\includegraphics[height=0.4\textheight]{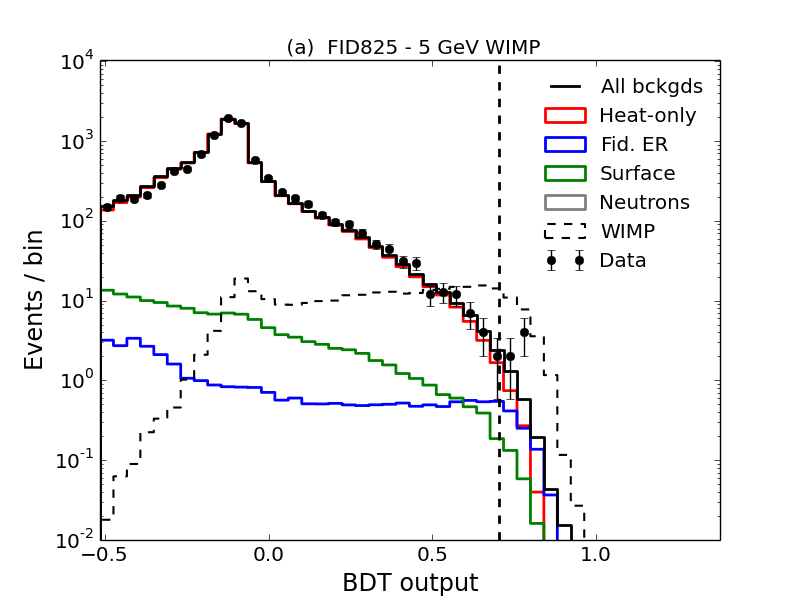}
\includegraphics[height=0.4\textheight]{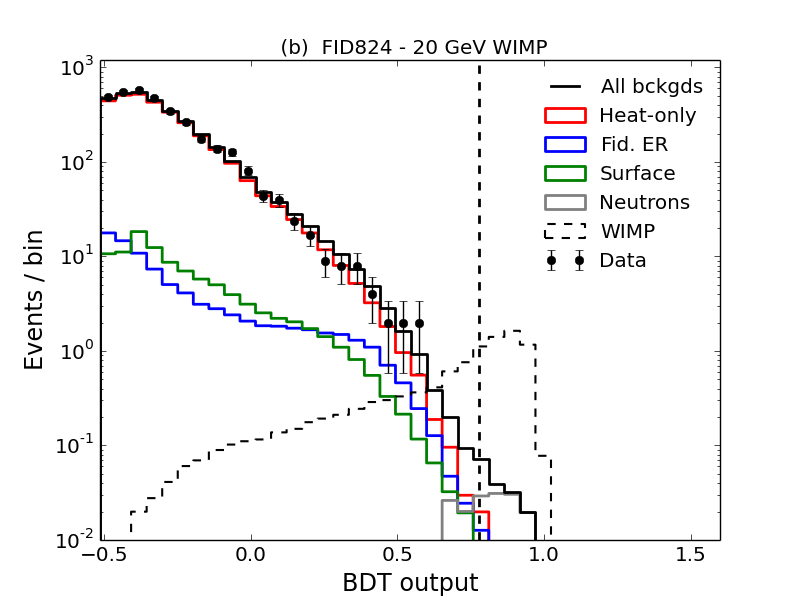}
\caption{Examples of BDT distributions for a $5\, {\rm GeV}/c^2$ (a) and $20 \,{\rm GeV}/c^2$ (b) WIMP mass. Continuous lines show the distributions for the different background models. The expected WIMP signal is represented with dashed lines, with an arbitrary normalization. The BDT distribution of the data in the RoI is overlaid together with the associated BDT cut (vertical dashed lines, see text).}
\label{fig:bdtdist}
\end{figure}

For each detector, a large amount of WIMP signal and background events are simulated within this RoI. Each simulated event consists in a set of 4 ionisation energies, one combined heat signal and a time-related variable which allows to connect to the rate of heat-only events at the time of the event. The time-varying noise of individual channels as well as online trigger efficiency are taken into account. All background event populations are extracted according to the models described in Section~2. WIMP-induced events within the fiducial volume are simulated assuming a NR spectrum as derived from~\cite{savage} with standard halo parameters. We use the NR quenching parametrization $Q_{\rm NR}=\alpha\,E_{\rm R}^{\beta}$ of Ref.~\cite{edw1}, where $E_{\rm R}$ is the recoil energy in keV, $\alpha=0.16$ and $\beta=0.18$, as it provides a good description of the AmBe neutron calibration data of the detectors. Both simulated signal and background events are used to train a Boosted Decision Tree (BDT) algorithm, which allows to reduce these different features into a single discriminating variable~\cite{these-thibault} normalized between $-1$ (background-like) and $+1$ (signal-like). A distinct BDT is trained for each of the eight detectors, and each WIMP mass among 10 tabulated masses in the $4-30$~GeV range.

The predicted BDT distributions of signal and backgrounds for two distinct masses and detectors are shown in Fig.~\ref{fig:bdtdist}. For each WIMP mass, the cuts on the BDT output of each detector are selected according to the values that optimize their combined sensitivity in the simulations, based on Poisson counting as described in~\cite{thesis-anderson}. The expected number of background events resulting from these cuts are listed in Table~\ref{tbl:unblinding}, together with their systematics as estimated by propagating the numbers quoted in Sect.~\ref{sec:bg}.

\begin{table}[!ht]
\centering
\renewcommand{\arraystretch}{1.5}
\setlength{\tabcolsep}{0.2cm}
\begin{tabular}{|l|c|c|c|c|}
\hline
WIMP mass & $5\, {\rm GeV}/c^2$ & $7\,{\rm GeV}/c^2$ & 10~${\rm GeV}/c^2$ & 20~${\rm GeV}/c^2$ \\
\hline\hline
Fiducial neutrons & $0.02\pm0.01$ & $0.15\pm0.07$ & $0.36\pm0.16$ & $1.05\pm 0.47$ \\
Fiducial ER & $2.71\pm 0.43$ & $1.02\pm0.16$ & $0.43\pm 0.07$ & $0.12\pm 0.02$ \\
Heat-only events & $2.87^{+0.49}_{-0.03}$ & $0.43^{+0.07}_{-0.00}$ & $0.20^{+0.03}_{-0.00}$ & $0.11^{+0.02}_{-0.00}$ \\
Others & $0.55\pm 0.16$ & $0.12\pm0.04$ & $0.09\pm 0.03$ & $0.07\pm 0.02$ \\
\hline
Total background & $6.14^{+0.67}_{-0.46}$ & $1.71^{+0.19}_{-0.18}$ & $1.07\pm0.18$ & $1.35\pm0.47$ \\
\hline\hline
Events observed & 9 & 6 & 4 & 4 \\
\hline
p-value & 22\% & 1.1\% & 2.8\% & 6.3\% \\
\hline
\end{tabular}
\caption{Expected and observed event statistics with their $1\sigma$ uncertainties for this WIMP search, after BDT cut and detector combination for four among the ten WIMP masses considered in this study. "Other" backgrounds include all non-fiducial events that were considered (beta, lead and gamma-rays with different ionisation pattern topologies).}
\label{tbl:unblinding}
\end{table}

\begin{figure}[!ht]
\centering
\includegraphics[height=0.27\textheight]{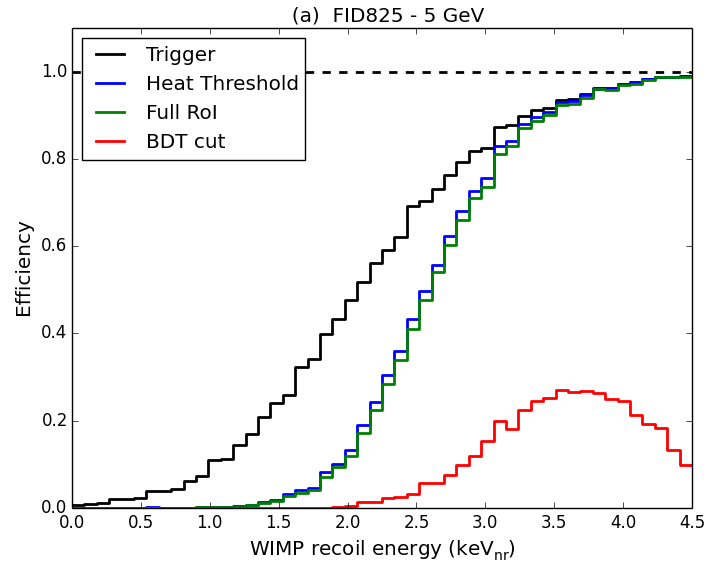}
\includegraphics[height=0.27\textheight]{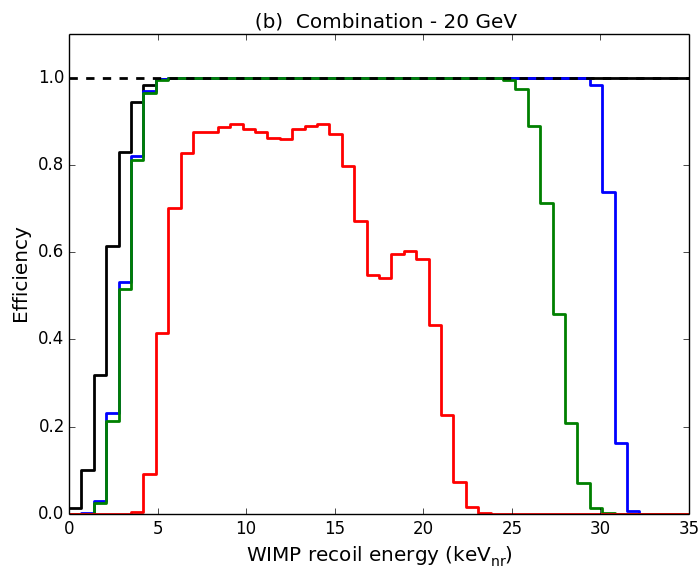}
\caption{Efficiency functions of different cuts for WIMP-induced nuclear recoils, as a function of the true, physical recoil energy. (a): $M_X=5 \,{\rm GeV}/c^2$, from the leading-sensitivity detector. (b): $M_X=20 \,{\rm GeV}/c^2$, from the detector combination. Black -- online trigger; blue -- analysis heat threshold; green -- Region of Interest; red -- BDT cut.}
\label{fig:efficiency}
\end{figure}

\begin{figure}[!ht]
\centering
\includegraphics[width=0.8\textwidth]{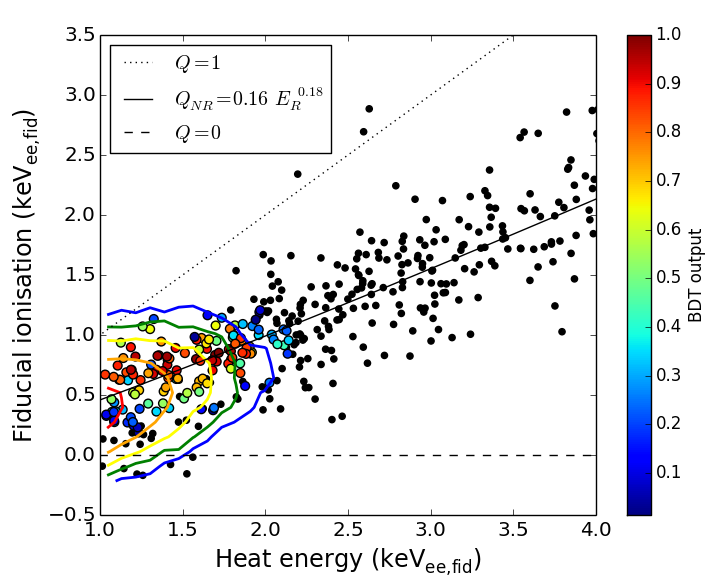}
\caption{Distribution in the heat vs fiducial ionisation plane of low-energy events recorded by FID825 during neutron calibration, after all cuts and processings identical to those applied for this WIMP search (black dots). The corresponding BDT scores for a $5 \,{\rm GeV}/c^2$ WIMP are illustrated when this score is positive. Colored contours represent the simulated distribution of $5 \,{\rm GeV}/c^2$ WIMP-induced recoils as expected in WIMP-search conditions. The solid, dotted and dashed lines represent the average ionisation yields for NR ($Q_{\rm NR}$), ER ($Q=1$) and heat-only events ($Q=0$) respectively.}
\label{fig:neutrons}
\end{figure}

The efficiency of successive cuts on the expected WIMP-induced NRs is illustrated in Fig.~\ref{fig:efficiency}.  For an event selection tuned to $M_X = 20\, {\rm GeV}/c^2$, the BDT keeps mostly recoils in the 6 to 15~keV (true NR) energy range. The former value corresponds to the threshold for the efficient rejection of gamma and heat-only backgrounds, while the latter corresponds to the rejection of the harder spectrum associated to the neutron background. The efficiency in the quoted range is 87\%. For a selection tuned to $M_X=5 \,{\rm GeV}/c^2$, the efficiency loss below 3~keV (NR scale) is mostly driven by the heat analysis threshold. Due to the soft recoil spectrum associated to low WIMP masses, only detectors with the 1~keV$_{\rm ee,fid}$ heat analysis threshold yield a significant efficiency. The stringent BDT cut then keeps only a fifth of all recoils in the $3-4$~keV (NR) range. The selection criteria derived by the BDT at low mass are better understood by studying its response to the neutron-induced nuclear recoils observed by exposing the detector to an AmBe source,  illustrated in Fig.~\ref{fig:neutrons}. Only the lowest-energy NR events acquire high BDT scores. In addition, the BDT favors events slightly above the average $Q_{\rm NR}$ curve. Indeed, while nuclear recoils lie at roughly equal distance from the heat-only ($Q = 0$) and fiducial ER ($Q = 1$), in WIMP-search data the intensity of the former exceeds largely that of the ER that are mostly due to the L-shell cosmic activation lines.  Therefore, the BDT naturally shifts the WIMP-search region towards slightly higher ionisation yields.

An additional consequence of the discrimination against the heat-only background is that most of the efficiency at low WIMP mass comes from the detector with the best ionisation resolution (FID825). This contrasts with the selection cuts tuned to $M_X=20 \,{\rm GeV}/c^2$, which result in a balanced share of sensitivity between all eight detectors.

\subsection{Results}

\begin{figure}[!ht]
\centering
\includegraphics[height=0.4\textheight]{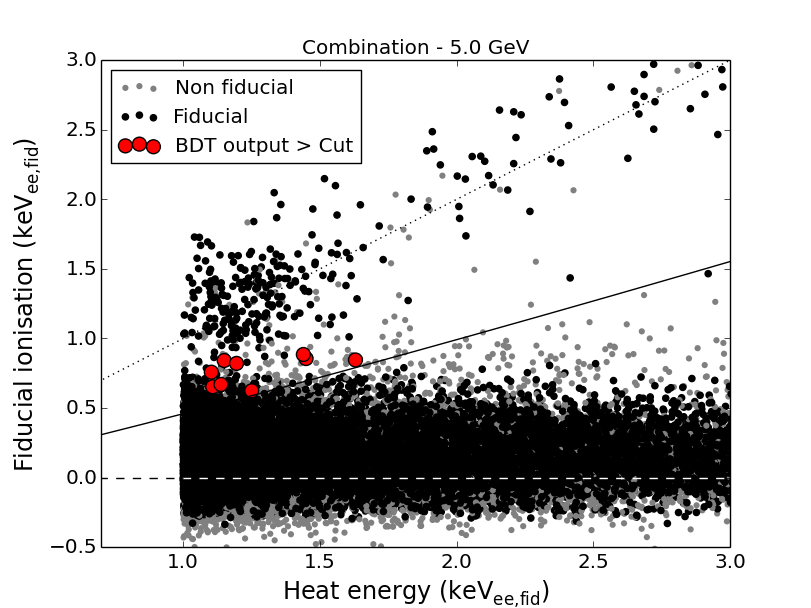}
\includegraphics[height=0.4\textheight]{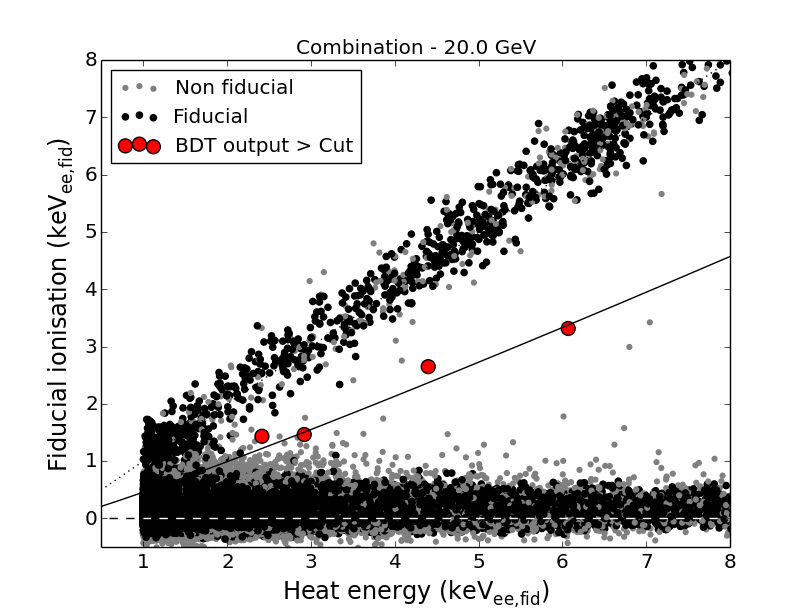}
\caption{Heat vs fiducial ionisation distributions for events in the RoI in the combined WIMP search for $M_X=5 \,{\rm GeV}/c^2$ (top) and $M_X=20 \,{\rm GeV}/c^2$ (bottom). Events with a significant veto signal ($>0.4$~keV$_{\rm ee,fid}$) are represented in grey. Events that pass the BDT cut are highlighted in red. The solid, dotted and dashed lines are as in Fig.~\ref{fig:neutrons}.}
\label{fig:scatterwimps}
\end{figure}

The event distributions after unblinding are illustrated in Fig.~\ref{fig:bdtdist} in terms of BDT output, and in Fig.~\ref{fig:scatterwimps} in the heat vs fiducial ionisation plane. Both the very high-statistics heat-only event population and L-shell lines between 1.1 and 1.3~keV are visible on Fig.~\ref{fig:scatterwimps}, with WIMP candidates in-between. Table~\ref{tbl:unblinding} compares the expected background with the observed event counts after the BDT cut for 5, 7, 10 and 20~${\rm GeV}/c^2$ WIMPs, and provides the corresponding p-values, defined as the probability to have more events than observed in the background-only hypothesis, given the Poisson statistics and gaussian systematic errors. While Fig.~\ref{fig:bdtdist} shows that the overall agreement between simulations and data in terms of BDT distributions is excellent, we observe slight excesses of events after the BDT cut for all WIMP masses. These excesses are associated to different events depending on the WIMP mass. For $M_X=20\,{\rm GeV}/c^2$, the four candidates come from three different detectors. They lie close to the $Q_{\rm NR}$ line, as shown in Fig.~\ref{fig:scatterwimps} (bottom). For $M_X=5\,{\rm GeV}/c^2$, 9 events are left as WIMP candidates, all of them in the $1-1.7$~keV$_{\rm ee,fid}$ heat energy range. The strongest excess in the whole $4-30\, {\rm GeV}/c^2$ mass range is for $M_X=7\,{\rm GeV}/c^2$, for which WIMP candidates from both the 5 and $20\,{\rm GeV}/c^2$ WIMP searches combine to yield a 1.1\% p-value.

\begin{figure}[!ht]
\centering
\includegraphics[width=0.9\textwidth]{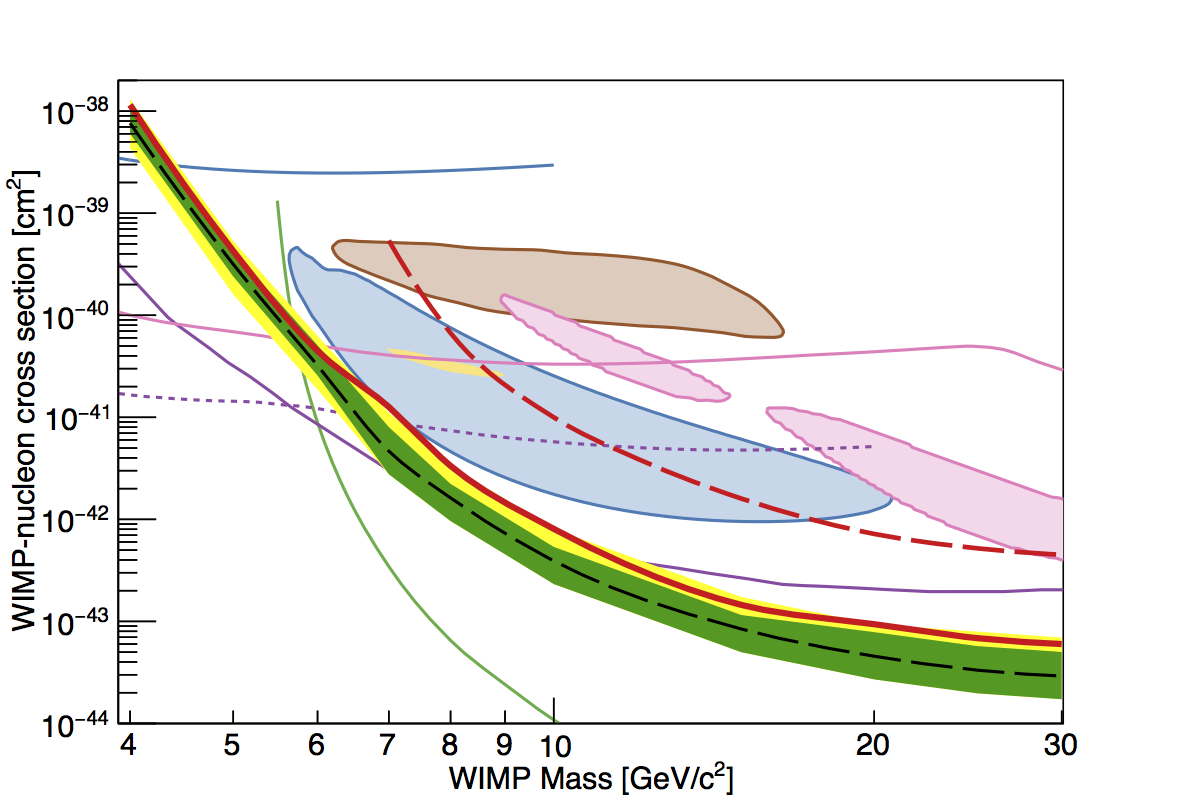}
\caption{Red curve: 90\% CL limit on the spin-independent WIMP-nucleon cross-section obtained in this work. The green (resp. yellow) band represents the expected $1\sigma$ (resp. $2\sigma$) sensitivity region in the absence of a signal. The yellow, blue, pink and brown contours are respectively from CoGeNT~\cite{cogent}, CDMS-Si~\cite{cdms-si}, CRESST-II~\cite{cresst} and DAMA~\cite{dama}. We also represent other limits by EDELWEISS-II~\cite{edw2-lowmass} (dashed red), LUX~\cite{lux} (green), DAMIC~\cite{damic} (blue), CRESST~\cite{cresst-light} (pink), CDMSLite~\cite{cdmslite} (dashed violet) and SuperCDMS~\cite{scdms} (violet).}
\label{fig:limit}
\end{figure}

We set a Poisson limit based only on the observed event count, such that the derived limit does not rely on any form of background subtraction. The corresponding 90\% CL limit on the spin-independent WIMP-nucleon cross-section is shown in Fig.~\ref{fig:limit} as a function of $M_X$, together with the expected sensitivity and its $1\,\sigma$ and $2\,\sigma$ uncertainty bands related to Poisson statistics and background systematics.

The following effects could affect the expected WIMP signal, and therefore the resulting limit. First, WIMP-induced interactions taking place outside the fiducial volume could enhance the low-mass WIMP sensitivity, since the fiducial volume selection rejects fewer surface events as ionisation signals become comparable to the ionisation resolution. Dedicated simulations demonstrate that the sensitivity gain is of the order of $1-2$\%. Systematics in our trigger efficiency model may affect the WIMP sensitivity by up to 5\% for $M_X=5\,{\rm GeV}/c^2$. Finally, since our analysis threshold is calibrated in keV$_{\rm ee,fid}$ and the WIMP search region for low masses is biased towards high quenching factors, uncertainties on the NR quenching factor are expected to impact the WIMP sensitivity especially for low WIMP masses. Neutron calibration data taken with FID detectors are compatible with our reference parametrization $Q_{\rm NR}=0.16\,E_{\rm R}^{0.18}$ over the whole energy range of interest, with an estimated uncertainty at the level of 5\%. Simulations demonstrate that a $\pm 5$~\% change in the quenching implies a $\pm 20$~\% change in the sensitivity to 5~${\rm GeV}/c^2$ WIMPs, and $\pm 4$~\% for $M_X>10\,{\rm GeV}/c^2$.

The excess of events above the expected background for $M_X>10\,{\rm GeV}/c^2$ is probably associated to  neutrons, since they are the dominant expected background in this case, with large uncertainty, and the observed events have ionisation yields very close to $Q_{\rm NR}$. However, for $M_X\leq 7\,{\rm GeV}/c^2$, three candidate events appear hardly compatible with the expected dominant backgrounds which consist of heat-only events and fiducial electron recoils. They are visible for  $M_X=5\,{\rm GeV}/c^2$ as the three highest-energy candidates on Fig.~\ref{fig:scatterwimps} (top). The post-unblinding investigations on their origin revealed a plausible additional source of background: calibrations with $^{210}$Pb sources on FID detectors suggest that up to 5\% of the $\beta$ background could appear as triple-electrode events, in which both fiducial electrodes are hit in addition to a veto electrode. Unlike triple-electrode events from the $\gamma$-ray background, they were not anticipated before unblinding. Depending on the assumed sharing of charge between the three electrodes, these $Q\sim0.4$ events may become undistinguishable from fiducial nuclear recoils at very low energy. Basic simulations support this scenario and confirm that the leaking events would have energies and fiducial ionisation yields similar to these three candidates. This would add up to 1.5 events to the expected background for the event selection tuned to $M_X=5\,{\rm GeV}/c^2$, resulting in a total expected background of 7.6 events and a p-value of 43\%. The lowest p-value, still for $M_X=7\,{\rm GeV}/c^2$, would then be 2.3\%.

\section{Conclusions and outlook}

In this article, the data recorded with an array of FID bolometers during a long background run at the LSM are used to set constraints on the scattering of $4-30\,{\rm GeV}/c^2$ mass dark matter particles with nucleons. During this first physics run for these newly developed detectors, they proved to be fully functional, both in terms of stability during long physics runs, and in terms of background rejection. Their remarkable capability for an ionisation-based fiducial selection at very low energies achieves very low levels of radioactive backgrounds down to 0.18 events per kg-day per keV within the fiducial volume of each detector. For the first time, all backgrounds relevant to the WIMP search were precisely modelled using data-driven parameterizations. These models were in turn used in order to optimize the WIMP-search region of experimental parameter space. In this study, a conservative constraint on the WIMP-nucleon cross-section is derived in a blind analysis, without any form of background subtraction. In a forthcoming publication~\cite{edw-likelihood}, we will use the full knowledge of our background in order to carry out a likelihood adjustment of our data, and better constrain a potential WIMP signal.

After unblinding, we found that the distribution of our data in the WIMP-search region was in reasonable agreement with the background models given their systematics. The corresponding p-values associated to event counting range from 1.1 to 22\%. Subsequently, as illustrated on Fig.~\ref{fig:limit}, spin-independent WIMP-nucleon scattering cross-sections are excluded at 90\% CL above $4.3\times 10^{-40}\,{\rm cm}^2$ (resp. $9.4\times 10^{-44}\,{\rm cm}^2$) for $M_X=5\,{\rm GeV}/c^2$ (resp. $M_X=20\,{\rm GeV}/c^2$).

The main sensitivity limitation of this search for WIMPs with $M_X<10\,{\rm GeV}/c^2$ is related to the presence of an intense background of heat-only events. These events prevent us from being sensitive to potential WIMP-induced recoils for which the deposited energy is large enough to be triggered by heat channels, but the quenching too small to generate a detectable fiducial ionisation signal. Consequently, a major priority of the experiment is to strongly reduce this background which is most probably of mechanical origin. Several tests are ongoing making use of different mechanical supports for the detectors. The reduction of the ionisation noise, potentially using HEMT amplifiers~\cite{hemt-hard,hemt-cdms}, will also alleviate this issue.

The results presented here constitute a major improvement with respect to the previous EDELWEISS-II low-mass WIMP search from~\cite{edw2-lowmass}. The derived limit is lower by a factor of 12 for $M_X=10\,{\rm GeV}/c^2$, and by a factor of 41 for $M_X=7\,{\rm GeV}/c^2$. This is related to an increased exposure and better baseline noise, especially for the ionisation channels, which enable to be sensitive to lower nuclear recoil energies. Our exclusion limit is in strong tension with a WIMP interpretation of the recent results in~\cite{dama,cogent,cresst,cdms-si}. Our result is therefore a complementary confirmation of the latest LUX~\cite{lux} and SuperCDMS~\cite{scdms} studies.

\section*{Acknowledgements}

The help of the technical staff of the Laboratoire Souterrain de Modane and the participant 
laboratories is gratefully acknowledged. The EDELWEISS project is supported in part by the 
German ministry of science and education (BMBF Verbundforschung ATP Proj.-Nr. 05A14VKA), 
by the Helmholtz Alliance for Astroparticle Phyics (HAP), by the French Agence Nationale 
pour la Recherche (ANR) and the LabEx Lyon Institute of Origins (ANR-10-LABX-0066) of 
the Universit\'e de Lyon within the program ``Investissements d'Avenir'' (ANR-11-IDEX-00007), 
by the P2IO LabEx (ANR-10-LABX-0038) in the framework ``Investissements d'Avenir'' 
(ANR-11-IDEX-0003-01) managed by the ANR (France), by Science and Technology Facilities 
Council (UK), and the Russian Foundation for Basic Research (grant No. 15-02-03561).


\begin{thebibliography}{99}
\bibitem{planck} Planck Collaboration, P. A.R. Ade et al., \emph{Planck 2013 results. XVI. Cosmological parameters}, \emph{Astron. Astrophys.} {\bf 571} (2014) A16 [\verb+arXiv:1303.5076+].
\bibitem{bertone} G. Bertone, D. Hooper and J. Silk, \emph{Particle dark matter: evidence, candidates and constraints}, \emph{Phys. Rept.} {\bf 405} (2005) 279-390 [\verb+arXiv:hep-ph/0404175+].
\bibitem{goodman} M. W. Goodman and E. Witten, \emph{Detectability of certain dark-matter candidates}, \emph{Phys. Rev.} {\bf D 31} (1985) 3059.
\bibitem{fermi} Fermi-LAT Collaboration, M. Ackermann et al., \emph{Searching for Dark Matter Annihilation from Milky Way Dwarf Spheroidal Galaxies with Six Years of Fermi Large Area Telescope Data}, \emph{Phys. Rev. Lett.} {\bf 115} (2015) 23, 231301 [\verb+arxiv:1503.02641+].
\bibitem{planck-2015} Planck Collaboration, P.A.R. Ade et al., \emph{Planck 2015 results. XIII. Cosmological parameters} [\verb+arxiv:1502.01589+].
\bibitem{kaplan} D. Kaplan, M. Luty and K. Zurek, \emph{Asymmetric dark matter}, \emph{Phys. Rev.} {\bf D 79} (2009) 115016 [\verb+arxiv:0901.4117+].
\bibitem{falkowski} A. Falkowski, J. Ruderman and T. Volanski, \emph{Asymmetric Dark Matter from Leptogenesis}, \emph{J. High Energy Phys.} {\bf 1105} (2011) 106 [\verb+arXiv:1101.4936+].
\bibitem{petraki} K. Petraki and R. Volkas, \emph{Review of asymmetric dark matter}, \emph{Int. J. Mod. Phys.} {\bf A28} (2013) 1330028 [\verb+arXiv:1305.4939+].
\bibitem{dama} DAMA Collaboration, R. Bernabei et al., \emph{Final model independent result of DAMA/LIBRA-phase1}, \emph{Eur. Phys. J.} {\bf C 73} (2013) 2648 [\verb+arXiv:1308.5109+].
\bibitem{cogent} CoGeNT Collaboration, C. Aalseth et al, \emph{Results from a Search for Light-Mass Dark Matter with a p-Type Point Contact Germanium Detector}, \emph{Phys. Rev. Lett.} {\bf 106} (2011) 131301 [\verb+arXiv:1002.4703+].
\bibitem{cresst} CRESST Collaboration, G. Angloher et al., \emph{Results from 730 kg days of the CRESST-II Dark Matter Search}, \emph{Eur. Phys. J.} {\bf C72} (2012) 1971 [\verb+arXiv:1109.0702+].
\bibitem{cdms-si} CDMS Collaboration, R. Agnese et al., \emph{Silicon Detector Dark Matter Results from the Final Exposure of CDMS II}, \emph{Phys. Rev. Lett.} {\bf 111} (2013) 25, 251301 [\verb+arXiv:1304.4279+].
\bibitem{kelko} C. Kelso, D. Hooper and M. Buckley, \emph{Toward a consistent picture for CRESST, CoGeNT, and DAMA}, \emph{Phys. Rev.} {\bf D85} (2012) 043515 [\verb+arXiv:1110.5338+].
\bibitem{kopp} J. Kopp, T. Schwetz and J. Zupan, \emph{Light Dark Matter in the light of CRESST-II}, \emph{JCAP} {\bf 1203} (2012) 001 [\verb+arXiv:1110.2721+].
\bibitem{lux} LUX Collaboration, D. Akerib et al., \emph{First Results from the LUX Dark Matter Experiment at the Sanford Underground Research Facility}, \emph{Phys. Rev. Lett.} {\bf 112} (2014) 091303 [\verb+arXiv:1310.8214+].
\bibitem{scdms} SuperCDMS Collaboration, R. Agnese et al., \emph{Search for Low-Mass Weakly Interacting Massive Particles with SuperCDMS}, \emph{Phys. Rev. Lett.} {\bf 112} (2014) 24, 241302 [\verb+arXiv:1402.7137+].
\bibitem{fid-gamma} A. Juillard (EDELWEISS Collaboration), \emph{Status and Prospects of the EDELWEISS Direct WIMP Search Experiment}, \emph{J. Low Temp. Phys.} {\bf 167} (2012) 1056.
\bibitem{edw2-lowmass} EDELWEISS Collaboration, E. Armengaud et al., \emph{Search for low-mass WIMPs with EDELWEISS-II heat-and-ionization detectors}, \emph{Phys. Rev.} {\bf D86} (2012) 051701(R) [\verb+arXiv:1207.1815+].
\bibitem{edw-etching} S. Marnieros et al. (EDELWEISS Collaboration), \emph{Controlling the Leakage-Current of Low Temperature Germanium Detectors Using XeF2 Dry Etching}, \emph{J. Low Temp. Phys.} {\bf 176} (2014) 182.
\bibitem{edw1} EDELWEISS Collaboration, V. Sanglard et al., \emph{Final results of the EDELWEISS-I dark matter search with cryogenic heat-and-ionization Ge detectors}, \emph{Phys. Rev.} {\bf D71} (2005)122002 [\verb+arXiv:astro-ph/0503265+].
\bibitem{edw2} EDELWEISS Collaboration, E. Armengaud et al., \emph{Final results of the EDELWEISS-II WIMP search using a 4-kg array of cryogenic germanium detectors with interleaved electrodes}, \emph{Phys. Lett. B} {\bf 702} (2011) 329 [\verb+arXiv:1103.4070+].
\bibitem{id} EDELWEISS Collaboration, A. Broniatowski et al., \emph{A new high-background-rejection dark matter Ge cryogenic detector}, \emph{Phys. Lett. B} {\bf 681} (2009) 305 [\verb+arXiv:0905.0753+].
\bibitem{fid-beta} J. Gascon and N. Bastidon (EDELWEISS Collaboration), \emph{The EDELWEISS-III Project and the Rejection Performance of Its Cryogenic Germanium Detectors}, \emph{J. Low Temp. Phys.} {\bf 176} (2014) 870.
\bibitem{edw-bg} EDELWEISS Collaboration, E. Armengaud et al., \emph{Background studies for the EDELWEISS dark matter experiment}, \emph{Astropart. Phys.} {\bf 47} (2013) 1 [\verb+arXiv:1305.3628+].
\bibitem{edw-setup} EDELWEISS Collaboration, \emph{Performance of the EDELWEISS-III array for the direct search of dark matter}, in preparation.
\bibitem{edw-bg-lrt} S. Scorza (EDELWEISS Collaboration), \emph{Background investigation in EDELWEISS-III}, \emph{AIP Conf. Proc.} {\bf 1672}, 100002 (2015).
\bibitem{edw-veto} EDELWEISS Collaboration, B. Schmidt et al., \emph{Muon-induced background in the EDELWEISS dark matter search}, \emph{Astropart. Phys.} {\bf 44} (2013) 28 [\verb+arXiv:1302.7112+].
\bibitem{edw-elec} EDELWEISS Collaboration, B. Censier et al., \emph{EDELWEISS Read-out Electronics and Future Prospects}, \emph{J. Low Temp. Phys.} {\bf 167} (2012) 645.
\bibitem{edw-crate} T. Bergmann et al., \emph{A Scalable DAQ System with High-Rate Channels and FPGA- and GPU-Trigger for the Dark Matter Search Experiment EDELWEISS-III}, IEEE NSS-MIC Conference Proceedings, San Diego, Nov 2015.
\bibitem{edw-thermal} J. Billard, M. De Jesus, A. Juillard and E. Queguiner (EDELWEISS Collaboration), \emph{Characterization and Optimization of EDELWEISS-III FID800 Heat Signals}, \emph{J. Low Temp. Phys.} (2016), doi:10.1007/s10909-016-1500-5.
\bibitem{edw-trapping} Q. Arnaud, EDELWEISS Collaboration, \emph{Signals Induced by Charge Carrier Trapping}, \emph{J. Low Temp. Phys.} {\bf 176} (2014) 924.
\bibitem{thesis-quentin} Q. Arnaud, PhD thesis, Universit\'{e} Claude Bernard Lyon 1 (2016) [\verb+https://tel.archives-ouvertes.fr/tel-01273303v1+].
\bibitem{thesis-cecile} C\'{e}cile K\'{e}f\'{e}lian, PhD thesis, Universit\'{e} Claude Bernard Lyon 1 and Karlsruhe Institute of Technology (2016).
\bibitem{edw-tritium} EDELWEISS Collaboration, \emph{Observation of tritium $\beta$ decay from cosmogenic activation in the volume of cryogenic germanium detectors}, in preparation.
\bibitem{bahcall} J. Bahcall, \emph{Exchange and Overlap Effects in Electron Capture and in Related Phenomena}, \emph{Phys. Rev.} {\bf 132}, 362 (1963).
\bibitem{cresst-cracks} J. Astrom et al., \emph{Fracture processes observed with a cryogenic detector}, \emph{Phys. Lett.} {\bf A356} (2006) 262-266 [\verb+arXiv:physics/0504151+].
\bibitem{savage} C. Savage, K. Freese and P. Gondolo, \emph{Annual Modulation of Dark Matter in the Presence of Streams}, \emph{Phys. Rev.} {\bf D74} (2006) 043531 [\verb+arxiv:astro-ph/0607121+].
\bibitem{these-thibault} T. Main de Boissi\`{e}re, PhD thesis, Universit\'{e} Paris Sud (2015) [\verb+https://tel.archives-ouvertes.fr/tel-01195586v1+].
\bibitem{thesis-anderson} A. Anderson, PhD thesis, Massachusetts Institute of Technology (2015).
\bibitem{damic} J. Barreto et al., \emph{Direct search for low mass dark matter particles with CCDs}, \emph{Phys. Lett. B} {\bf  711}  (2012) 264?269 [\verb+arXiv:1105.5191+].
\bibitem{cresst-light} CRESST Collaboration, G. Angloher et al., \emph{Results on light dark matter particles with a low-threshold CRESST-II detector}, \emph{Eur. Phys. J.} {\bf C 76} (2016) 25 [\verb+arXiv:1509.01515+].
\bibitem{cdmslite} SuperCDMS Collaboration, R. Agnese et al., \emph{New Results from the Search for Low-Mass Weakly Interacting Massive Particles with the CDMS Low Ionization Threshold Experiment}, \emph{Phys. Rev. Lett.} {\bf 116} (2016) 071301 [\verb+arXiv:1509.02448+].
\bibitem{edw-likelihood} EDELWEISS Collaboration, \emph{Improved EDELWEISS-III sensitivity for low mass WIMPs using a profile likelihood approach}, in preparation.
\bibitem{hemt-hard} X. de la Broise and A. Bounab, \emph{Cryogenic ultra-low noise HEMT amplifiers board}, \emph{Nucl. Instrum. Meth.} {\bf A787} (2015) 51.
\bibitem{hemt-cdms} A. Phipps, B. Sadoulet, A. Juillard and Y. Jin, \emph{An HEMT-Based Cryogenic Charge Amplifier for Sub-kelvin Semiconductor Radiation Detectors},  \emph{J. Low Temp. Phys.} (2016), http://dx.doi.org/10.1007/s10909-016-1475-2.
\end{thebibliography}
\end{document}